\documentclass[twocolumn,showpacs,floats,floatfix,superscriptaddress,aps,pra]{revtex4-1}
\usepackage{eurosym}
\usepackage{amsfonts}
\usepackage{amssymb}
\usepackage{amsmath}
\usepackage{graphicx}
\usepackage{bm}
\usepackage{color}
\usepackage{xcolor}
\usepackage{epsfig}
\usepackage{ifthen}
\usepackage{subfigure}
\usepackage{physics}
\usepackage{float}
\usepackage[hidelinks, colorlinks=true,linkcolor=blue,citecolor=blue,filecolor=blue,urlcolor=blue]{hyperref}

\begin{document}

\title{Robust population transfer by a detuning sign jump:\\
from two-state quantum system to SU(2)-symmetric three-state quantum system}
\author{Peter Chernev}
\affiliation{Center for Quantum Technologies, Department of Physics, Sofia University,
James Bourchier 5 blvd., 1164 Sofia, Bulgaria}
\author{Andon A. Rangelov}
\affiliation{Center for Quantum Technologies, Department of Physics, Sofia University,
James Bourchier 5 blvd., 1164 Sofia, Bulgaria}
\date{\today }

\begin{abstract}
We propose and analyze a robust population-transfer protocol in a driven
two-level system based on a sudden \emph{sign} change of the detuning at the
maximum of a smooth coupling pulse. Away from the jump the dynamics is
adiabatic, while the sign flip produces a single nonadiabatic ``kick'' in
the adiabatic basis. Within a simple stepwise adiabatic–sudden approximation we obtain a compact analytic expression for the final transition probability, identify the parameter regimes that yield high-fidelity inversion, and show that the result depends only on the change of the mixing angle across the detuning jump, i.e., solely on the ratio of the peak Rabi frequency to the detuning. Numerical simulations
of the full time-dependent Schr\"odinger equation confirm the validity and
robustness of this description over a broad parameter range.

We then use the Majorana decomposition to extend the scheme to an
SU(2)-symmetric three-state chain driven by the \emph{same} coupling and
detuning functions. In this setting the three-state propagator is expressed
in closed form through the two-level Cayley--Klein parameters, which allows
us to derive explicit transition probabilities for all three initial states.
In particular, we show that for strong coupling the protocol yields almost
complete population transfer between the two outer states, with only small
transient population of the middle state, while retaining the same intrinsic
robustness as in the underlying two-level model.
\end{abstract}


\maketitle

\section{Introduction}

\label{sec:intro}

Coherent control of two-state quantum systems is a central task in fields
ranging from atomic and molecular physics to quantum information, nonlinear
optics, and nuclear magnetic resonance\cite%
{Allen-Eberly,Shore,Yatsenko,Vitanov2001A,Vitanov2001B,Klein,Levitt79,Freeman80,Levitt86,Freeman97}%
. Among the many techniques developed for robust state manipulation, two
approaches stand out for their generality and wide applicability: rapid
adiabatic passage (RAP) \cite%
{Allen-Eberly,Shore,Yatsenko,Vitanov2001A,Vitanov2001B,Klein} and
composite-pulses (CPs) \cite{Levitt79,Freeman80,Levitt86,Freeman97}. RAP
achieves population inversion by slowly varying the parameters of the
driving field so that the system follows an instantaneous adiabatic
eigenstate from one diabatic state to the other. When implemented correctly,
RAP is highly robust against parameter fluctuations, but its reliance on
adiabaticity typically requires long interaction times or specially
engineered chirps to ensure fulfillment of adiabatic condition throughout
the pulse interaction. Composite pulses \cite%
{Levitt79,Freeman80,Levitt86,Freeman97}, by contrast, rely on sequences of
resonant or near-resonant rotations and can achieve robust transfer even
with short pulses, but at the cost of increased control complexity and
sensitivity to phase errors.

Alternative approaches achieve robustness through nonadiabatic but carefully
engineered pulses. Example is phase-jump schemes \cite{Vitanov2007,Torosov2007,Lehto,Zlatanov}, in which the Rabi frequency
changes sign or phase at the pulse maximum. In some of these models, the dynamics can
be solved exactly, revealing simple conditions for complete population
inversion and broad parameter regions of robustness.

In this work we propose and analyze a related but distinct protocol. We
assume that the Rabi frequency has a smooth pulse shape, such as a Gaussian, $\sech$, Lorentzian,  etc.
while the detuning is \emph{constant in magnitude} but undergoes a sudden
sign flip at the peak of the coupling. The evolution is adiabatic for times
well before and after the sign change, and the detuning jump produces a
single, localized nonadiabatic event in the adiabatic basis. We treat the
dynamics in an adiabatic--sudden approximation and derive a simple analytic
expression for the final population transfer. We then support this picture
by numerical simulations and show that the protocol is robust to moderate
variations of the pulse parameters.
We further extend the idea to an SU(2)-symmetric three-state system, in
which three equally spaced diabatic levels are coupled by the \emph{same}
time-dependent Rabi frequency and detuning as in the two-level model. Using
the Majorana decomposition \cite{Majorana,Bloch,Hioe}, the three-state
dynamics is mapped onto the underlying two-level problem, and the full
$3\times 3$ propagator is expressed in closed form in terms of the
two-state Cayley--Klein parameters \cite{Ramakrishna}. This construction allows us to transfer
the detuning-jump protocol directly to the three-state chain and to derive
explicit analytical expressions for all transition probabilities. In
particular, we show that for strong coupling the detuning sign jump induces
almost complete population transfer between the two outer states, with only
a small final population of the middle state, while preserving
the robustness inherited from the two-state dynamics.

\section{Two-level model with a detuning sign jump}
\label{sec:twolevel}

\subsection{Hamiltonian, protocol and basic quantities}

We consider a generic driven two-level quantum system with basis states $%
\{|1\rangle ,|2\rangle \}$, often referred to as the \emph{diabatic} or 
\emph{bare} states. These may represent, for example, two internal atomic
levels, two spin states, or any pair of states coupled by an external
classical field. In the rotating-wave approximation (RWA)\cite%
{Allen-Eberly,Shore}, and for a driving field with real phase, the
Hamiltonian takes the standard form 
\begin{equation}
H(t)=\frac{\hbar }{2} 
\begin{pmatrix}
-\Delta (t) & \Omega (t) \\[4pt] 
\Omega (t) & \Delta (t)%
\end{pmatrix}
,  \label{eq:H-diabatic}
\end{equation}
where $\Omega (t)$ is the (real) \emph{Rabi frequency}, proportional to the
coupling between the two levels, while $\Delta (t)$ is the \emph{detuning}
between the instantaneous driving frequency and the transition frequency
between $|1\rangle$ and $|2\rangle$. Both $\Omega(t)$ and $\Delta(t)$ may be
controlled externally via the amplitude and frequency of the driving field.

The state of the system is written as a superposition 
\begin{equation}
|\Psi (t)\rangle =c_{1}(t)|1\rangle +c_{2}(t)|2\rangle ,
\end{equation}
with probability amplitudes $c_{1}(t)$ and $c_{2}(t)$ in the diabatic basis.
Substituting this expansion into the time-dependent Schr\"odinger equation
yields 
\begin{equation}
i\hbar \frac{d}{dt} 
\begin{pmatrix}
c_{1} \\ 
c_{2}%
\end{pmatrix}
=\frac{\hbar }{2} 
\begin{pmatrix}
-\Delta & \Omega \\[4pt] 
\Omega & \Delta%
\end{pmatrix}
\begin{pmatrix}
c_{1} \\ 
c_{2}%
\end{pmatrix}
,  \label{Schroudinger equation}
\end{equation}
which fully determines the evolution.

We now take the coupling to be a smooth pulse shape of the form, 
\begin{equation}
\Omega (t)=\Omega _{0}f\left( t\right) ,  \label{eq:Omega-Gaussian}
\end{equation}%
where $\Omega _{0}$ is the peak Rabi frequency and function $f\left( t\right) $ is
smooth pulse shape such as Gaussian, $\sech$, Lorentzian,  etc. with a maximum at $t=0$.
The key ingredient of the protocol is the detuning profile. We assume the
detuning has constant magnitude $\Delta_0 > 0$ but undergoes a sign flip at $%
t=0$, 
\begin{equation}
\Delta(t) = 
\begin{cases}
+\Delta_0, & t < 0, \\ 
-\Delta_0, & t > 0 .%
\end{cases}
\label{eq:Delta-sign-jump}
\end{equation}
The sign change is assumed to occur on a time scale much shorter than $%
1/\varepsilon(0)$ (defined below), so that it can be treated as
instantaneous. The eigenenergy splitting of the two dressed states is 
\begin{equation}
\varepsilon (t)=\sqrt{\Omega ^{2}(t)+\Delta ^{2}(t)},
\end{equation}%
and the mixing angle 
\begin{equation}
\tan 2\theta (t)=\frac{\Omega (t)}{\Delta (t)},  \label{tita}
\end{equation}%
which characterizes the instantaneous rotation between diabatic and
adiabatic bases. These parameters play a central role in the analysis of
rapid adiabatic passage (RAP) \cite%
{Allen-Eberly,Shore,Yatsenko,Vitanov2001A,Vitanov2001B}.

We consider the system initially prepared in the ground state, 
\begin{equation}
\ket{\psi(t\to -\infty)}=\ket{1},
\end{equation}
and we are interested in the final population in $\ket{2}$ at $t\rightarrow
+\infty$. Our strategy is to assume adiabatic evolution before and after the
detuning jump, derive the corresponding evolution operators in the adiabatic
basis, include the instantaneous mixing at the jump, and finally transform
back to the diabatic basis to obtain the total propagator and the transition
probability.


\section{Adiabatic basis and equations of motion}

\subsection{Instantaneous eigenstates and basis transformation}

To analyze the evolution, we work in the \emph{adiabatic basis} \cite%
{Vitanov2001A,Vitanov2001B}, i.e.\ the instantaneous eigenbasis of the
Hamiltonian~\eqref{eq:H-diabatic}. Diagonalizing $H(t)$ yields two
time-dependent eigenenergies 
\begin{equation}
E_{\pm }(t)=\pm \frac{\hbar \varepsilon (t)}{2},
\end{equation}
which correspond to the upper and lower adiabatic energy surfaces.

The associated eigenvectors are naturally expressed in terms of the mixing
angle $\theta (t)$ defined in Eq.~\eqref{tita}, which describes how the
diabatic basis states $\{|1\rangle ,|2\rangle \}$ mix due to the applied
field. In analogy with standard RAP conventions \cite%
{Vitanov2001A,Vitanov2001B}, we choose real normalized eigenstates, 
\begin{equation}
|+(t)\rangle = 
\begin{pmatrix}
\cos \theta \\[3pt] 
\sin \theta%
\end{pmatrix}
, \qquad |-(t)\rangle = 
\begin{pmatrix}
-\sin \theta \\[3pt] 
\cos \theta%
\end{pmatrix}
.  \label{eigenvectors}
\end{equation}
As the external fields vary, the mixing
angle $\theta (t)$ changes, causing these eigenvectors to rotate relative to
the diabatic basis.

Any quantum state may be expanded in this instantaneous adiabatic basis as 
\begin{equation}
|\Psi(t)\rangle = a_+(t)|+(t)\rangle + a_-(t)|-(t)\rangle,
\end{equation}
where $a_+(t)$ and $a_-(t)$ are the \emph{adiabatic amplitudes}, describing
the populations of the eigenstates during the evolution.

The relation between the diabatic amplitudes $\mathbf{c}(t)=(c_{1},c_{2})^{T}
$ and the adiabatic amplitudes $\mathbf{a}(t)=(a_{+},a_{-})^{T}$ is a simple
rotation, implemented by the matrix composed of the eigenvectors~%
\eqref{eigenvectors}, 
\begin{equation}
R(t)= 
\begin{pmatrix}
\cos {\theta } & -\sin {\theta } \\[4pt] 
\sin {\theta } & \cos {\theta }%
\end{pmatrix}
.  \label{rotation}
\end{equation}
This gives 
\begin{equation}
\mathbf{c}(t)=R(t)\,\mathbf{a}(t),\qquad \mathbf{a}(t)=R^{T}(t)\,\mathbf{c}%
(t),
\end{equation}
where $R^{T}$ denotes the transpose (inverse) matrix of $R$. The time dependence of $R(t)$
encodes the changing geometry of the adiabatic basis and is, as we now show,
the source of nonadiabatic couplings whenever $\dot{\theta}(t)\neq 0$.

\subsection{Adiabatic-frame Schr\"odinger equation and adiabaticity}

Because the adiabatic states themselves depend on time through the mixing
angle $\theta(t)$, transforming the Hamiltonian into this basis requires us
to account both for the diagonalization of $H(t)$ and for the explicit time
dependence of the basis vectors. The adiabatic-frame Hamiltonian is obtained
from the diabatic one by the unitary transformation 
\begin{equation}
H_{\mathrm{ad}}(t) = R^T(t)\, H(t)\, R(t) - i\hbar\, R^T(t)\, \dot R(t),
\end{equation}
where the first term rotates the Hamiltonian into the instantaneous
eigenbasis, while the second term arises from the time derivative of the
basis itself. The additional term is responsible for nonadiabatic
transitions between the instantaneous eigenstates whenever the mixing angle
changes in time.

A straightforward calculation gives 
\begin{equation}
H_{\mathrm{ad}}(t)=\frac{\hbar }{2} 
\begin{pmatrix}
-\varepsilon (t) & -i\dot{\theta}(t) \\[6pt] 
i\dot{\theta}(t) & \varepsilon (t)%
\end{pmatrix}
.
\end{equation}
The diagonal elements $\pm \varepsilon (t)/2$ are the adiabatic
eigenenergies. The off-diagonal elements, proportional to $\dot{\theta}(t)$,
describe the \emph{nonadiabatic coupling} between the adiabatic states:
whenever $\dot{\theta}(t)$ is nonzero, population can leak from one
adiabatic state to the other.

In terms of the adiabatic amplitudes $\mathbf{a}(t) = (a_+, a_-)^T$, the
Schr\"odinger equation becomes 
\begin{equation}
i\frac{d}{dt} 
\begin{pmatrix}
a_+ \\ 
a_-%
\end{pmatrix}
= \frac{1}{2} 
\begin{pmatrix}
-\varepsilon & -i\dot{\theta} \\[6pt] 
i\dot{\theta} & \varepsilon%
\end{pmatrix}
\begin{pmatrix}
a_+ \\ 
a_-%
\end{pmatrix}
.
\end{equation}
Starting from the definition of the mixing angle, 
\begin{equation}
\tan 2\theta(t)=\frac{\Omega(t)}{\Delta(t)},
\end{equation}
or $2\theta(t) = \arctan[\Omega(t)/\Delta(t)]$, direct differentiation gives 
\begin{equation}
2\dot{\theta}(t) = \frac{\Delta\,\dot{\Omega} - \Omega\,\dot{\Delta}} {%
\Omega^2 + \Delta^2},
\end{equation}
which makes it clear that nonadiabatic effects arise only when the ratio $%
\Omega(t)/\Delta(t)$ changes in time.

The magnitude of the nonadiabatic coupling must be small compared to the
instantaneous eigenenergy splitting in order for adiabatic following to
hold. This yields the familiar adiabaticity condition 
\begin{equation}
|\dot{\theta}(t)|\ll \varepsilon (t),  \label{eq:adiabaticity-inequality}
\end{equation}%
which we will assume to be satisfied for $t<0$ and for $t>0$ in our step
adiabatic treatment.


\section{Step adiabatic evolution and total propagator}
\label{sec:Step adiabatic evolution and total propagator}

\subsection{Adiabatic evolution before and after the jump}

In the strictly adiabatic regime, the nonadiabatic coupling $\dot{\theta}(t)$
is negligibly small and the off-diagonal terms in $H_{\mathrm{ad}}(t)$ can
be ignored; the system then remains in a single adiabatic eigenstate,
acquiring only a dynamical phase. In our protocol, this approximation is
applied \emph{stepwise}: it is assumed to hold for $t<0$ and for $t>0$,
while the instantaneous detuning sign flip at $t=0$ produces mixing between
the adiabatic states.

We therefore view the dynamics as three consecutive steps:

\begin{enumerate}
\item Adiabatic evolution from an initial time $t_i \to -\infty$ up to  $t =
0^-$.

\item An instantaneous basis rotation at the detuning jump, connecting the 
adiabatic eigenstates just before and just after the jump.

\item Adiabatic evolution from $t = 0^+$ to a final time $t_f \to +\infty$.
\end{enumerate}

In the intervals $t<0$ and $t>0$ the equations of motion reduce to 
\begin{align}
i\dot{a}_{+}^{(<)}(t)& =-\tfrac{1}{2}\varepsilon (t)\,a_{+}^{(<)}(t), & i%
\dot{a}_{+}^{(>)}(t)& =-\tfrac{1}{2}\varepsilon (t)\,a_{+}^{(>)}(t), \\[4pt]
i\dot{a}_{-}^{(<)}(t)& =\phantom{-}\tfrac{1}{2}\varepsilon
(t)\,a_{-}^{(<)}(t), & i\dot{a}_{-}^{(>)}(t)& =\phantom{-}\tfrac{1}{2}%
\varepsilon (t)\,a_{-}^{(>)}(t),
\end{align}%
where $\varepsilon (t)$ is computed with $\Delta (t)=+\Delta _{0}$ for $t<0$
and with $\Delta (t)=-\Delta _{0}$ for $t>0$. Integrating from $t_{i}$ to $%
0^{-}$ and from $0^{+}$ to $t_{f}$ we obtain 
\begin{widetext}
\begin{align}
  a_+^{(<)}(0^-) &= a_+^{(<)}(t_i)\exp\!\left(
                      i\frac{\alpha_-}{2}\right), &
  a_+^{(>)}(t_f) &= a_+^{(>)}(0^+)\exp\!\left(
                      i\frac{\alpha_+}{2}\right), \\[4pt]
  a_-^{(<)}(0^-) &= a_-^{(<)}(t_i)\exp\!\left(
                     -i\frac{\alpha_-}{2}\right), &
  a_-^{(>)}(t_f) &= a_-^{(>)}(0^+)\exp\!\left(
                     -i\frac{\alpha_+}{2}\right),
\end{align}
\end{widetext} with the accumulated phases 
\begin{equation}
\alpha _{-}=\int_{t_{i}}^{0}\varepsilon (t^{\prime })\,dt^{\prime },\qquad
\alpha _{+}=\int_{0}^{t_{f}}\varepsilon (t^{\prime })\,dt^{\prime }.
\end{equation}%
It is convenient to define 
\begin{equation}
\delta _{-}=\frac{\alpha _{-}}{2},\qquad \delta _{+}=\frac{\alpha _{+}}{2},
\end{equation}%
so that the adiabatic evolution for $t<0$ and $t>0$ is represented simply by
the diagonal phase factors $\exp (\pm i\delta _{-})$ and $\exp (\pm i\delta
_{+})$.

\subsection{Detuning jump and total propagator}

The sudden flip of the detuning at $t=0$ changes $\Delta$ from $+\Delta_0$
to $-\Delta_0$, and therefore also the mixing angle $\theta(t)$ and the
corresponding adiabatic eigenstates. We denote 
\begin{equation}
\theta_- = \theta(0^-), \qquad \theta_+ = \theta(0^+),
\end{equation}
and define their difference 
\begin{equation}
\delta\theta = \theta_- - \theta_+ .
\end{equation}
The adiabatic eigenvectors just before and just after the jump are related
by a rotation, 
\begin{equation}
\begin{pmatrix}
|+(0^-)\rangle \\[3pt] 
|-(0^-)\rangle%
\end{pmatrix}
= 
\begin{pmatrix}
\cos\delta\theta & -\sin\delta\theta \\[4pt] 
\sin\delta\theta & \phantom{-}\cos\delta\theta%
\end{pmatrix}
\begin{pmatrix}
|+(0^+)\rangle \\[3pt] 
|-(0^+)\rangle%
\end{pmatrix}%
.  \label{eq:jump-rotation}
\end{equation}
Because the jump is assumed to be much faster than $1/\varepsilon(0)$, we
neglect any phase accumulation during the jump itself; the matrix in Eq.~%
\eqref{eq:jump-rotation} then represents the entire effect of the detuning
flip on the adiabatic amplitudes.

In the adiabatic basis, the full propagator from $t_{i}$ to $t_{f}$ is the
product of the three contributions: evolution for $t<0$, the instantaneous
rotation at $t=0$, and evolution for $t>0$. In matrix form, 
\begin{widetext}
\begin{equation}
  U_{\mathrm{ad}}(t_f,t_i)
  =
  \underbrace{
  \begin{pmatrix}
    e^{i\delta_+} & 0 \\[3pt]
    0             & e^{-i\delta_+}
  \end{pmatrix}}_{\text{adiabatic evolution for }t>0}
  \underbrace{
  \begin{pmatrix}
    \cos\delta\theta & -\sin\delta\theta \\[4pt]
    \sin\delta\theta & \phantom{-}\cos\delta\theta
  \end{pmatrix}}_{\text{detuning jump at }t=0}
  \underbrace{
  \begin{pmatrix}
    e^{i\delta_-} & 0 \\[3pt]
    0             & e^{-i\delta_-}
  \end{pmatrix}}_{\text{adiabatic evolution for }t<0}.
  \label{eq:Uad-piecewise}
\end{equation}
\end{widetext} The three factors in Eq.~\eqref{eq:Uad-piecewise} act
successively on the adiabatic amplitudes $(a_{+},a_{-})^{T}$.

To obtain the propagator in the physical diabatic basis $\{|1\rangle,|2%
\rangle\}$, we use the rotation matrix $R(t)$ [Eq.~\eqref{rotation}], which
relates the two bases at the initial and final times: 
\begin{equation}
U(t_f,t_i) = R(t_f)\,U_{\mathrm{ad}}(t_f,t_i)\,R^T(t_i).
\label{eq:U-diabatic-general}
\end{equation}
For our protocol we take $t_i\to-\infty$ and $t_f\to+\infty$. In these
limits the coupling vanishes, $\Omega(t_i,t_f)\to 0$, and the mixing angle tends
to fixed values determined solely by the asymptotic detuning. For $t\to
-\infty$ we have $\Delta(t)\to +\Delta_0$, so that $\theta(-\infty)=0$ and thus
\begin{equation}
R(t_i\to -\infty) = 
\begin{pmatrix}
1 & 0 \\[2pt] 
0 & 1%
\end{pmatrix}%
.
\label{eq:asymptotic1}
\end{equation}
For $t\to +\infty$ we have $\Delta(t)\to -\Delta_0$, which implies $%
\theta(+\infty)=\pi/2$ and thus 
\begin{equation}
R(t_f\to +\infty) = 
\begin{pmatrix}
0 & -1 \\[2pt] 
1 & \phantom{-}0%
\end{pmatrix}%
.
\label{eq:asymptotic2}
\end{equation}

The final propagator is conveniently parameterized with the complex Cayley--Klein
parameters \cite{Ramakrishna} $a$ and $b$ as 
\begin{equation}
U(+\infty,-\infty)=
\begin{pmatrix}
a & b \\[4pt]
-b^{\ast } & a^{\ast }
\end{pmatrix},
\qquad |a|^{2}+|b|^{2}=1,
\label{U}
\end{equation}
and comparison with Eq.~\eqref{eq:Uad-piecewise} together with the
asymptotic rotations $R(t_i)$ and $R(t_f)$ shows that in our case
\begin{align}
a &= -\,e^{i(\delta_{-}-\delta_{+})}\,\sin\delta\theta, 
\label{eq:a-CK}\\[4pt]
b &= -\,e^{-i(\delta_{-}+\delta_{+})}\,\cos\delta\theta .
\label{eq:b-CK}
\end{align}
Of particular interest is the element $U_{21}$, which gives the amplitude
to start in $\ket{1}$ and end in $\ket{2}$. From Eq.~\eqref{U} one has
\begin{equation}
U_{21} = -\,b^{\ast} 
= e^{i(\delta_{-}+\delta_{+})}\,\cos\delta\theta,
\end{equation}
so that the final population in $\ket{2}$ is 
\begin{equation}
P_2(\infty) = |U_{21}|^2 = \cos^2\delta\theta
= \cos^2\!\bigl(\theta_- - \theta_+\bigr).
\label{eq:P2-final-cos2}
\end{equation}
Thus, in this step adiabatic picture the inversion fidelity is determined
solely by the change in mixing angle across the detuning jump. In
our symmetric case $\Delta(t)=\pm\Delta_0$ one finds 
$\theta_+ = \tfrac{\pi}{2}-\theta_-$, and Eq.~\eqref{eq:P2-final-cos2}
reduces to
\begin{equation}
P_2(\infty) = \frac{\Omega_0^2}{\Omega_0^2 + \Delta_0^2},
\label{eq:P2-final-frac}
\end{equation}
which depends only on the ratio $\Omega_0/\Delta_0$. Deviations from perfect
adiabaticity before and after the jump, as well as a finite jump duration,
will modify Eq.~\eqref{eq:P2-final-cos2}, but, as shown in the next section, the basic
dependence on $\theta_- - \theta_+$ (and thus on $\Omega_0$ and $\Delta_0$)
remains robust over a broad parameter range.

\section{Numerical validation and robustness}
\label{sec:numerics}

To test the accuracy of the adiabatic--sudden approximation and the
robustness of the detuning--jump protocol, we numerically solve the
time-dependent Schr\"odinger equation with the Hamiltonian
\eqref{eq:H-diabatic}, using the coupling profile \eqref{eq:Omega-Gaussian}
and the detuning profile \eqref{eq:Delta-sign-jump}. We then compare the
final excited-state population $P_{2}(\infty)$ with the analytic prediction
\eqref{eq:P2-final-frac}. 

For the coupling we take a Gaussian pulse shape
\begin{equation}
f(t) = \exp\!\left(-\frac{t^{2}}{2T^{2}}\right),
\end{equation}
where $T$ is the pulse width. Time is measured in units of $T$, and the
parameters $\Omega_{0}$ and $\Delta_{0}$ are expressed in units of $1/T$.
\begin{figure}[htb]
\includegraphics[width=\columnwidth]{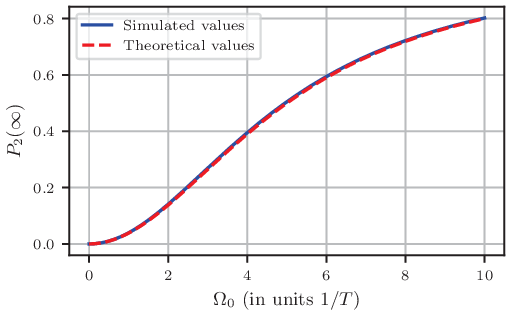}
\caption{(Color online) Final transition probability $P_2(\infty)$ as a
function of the peak Rabi frequency $\Omega_0$ for fixed detuning
$\Delta_0 = 5/T$. The solid blue curve shows the numerical result obtained
from integrating Eq.~\eqref{Schroudinger equation}, while the red dotted
curve shows the analytical approximation~\eqref{eq:P2-final-frac}.}
\label{fig1}
\end{figure}

The numerical integration is performed over the interval
$t \in [-20T,20T]$, which is sufficiently long that the coupling is
effectively zero at the boundaries and the asymptotic conditions of
Sec.~\ref{sec:Step adiabatic evolution and total propagator} are well satisfied. We compute $P_{2}(\infty)$ on a
two-dimensional grid of parameters with
$0 < \Omega_{0}T \leq 10$ and $0 < \Delta_{0}T \leq 10$, and compare the
resulting transition probabilities to the analytic formula
\eqref{eq:P2-final-frac}.

\begin{figure}[htb]
\includegraphics[width=\columnwidth]{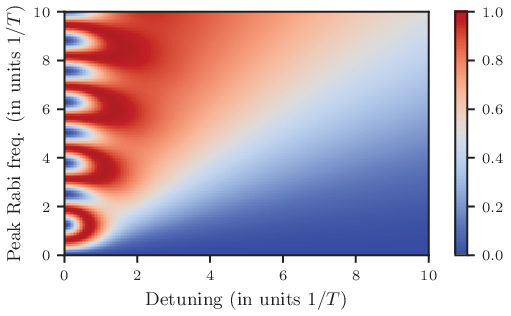}
\caption{(Color online) Contour plot of the final transition probability
$P_2(\infty)$ obtained from numerical integration of the Schr\"odinger
equation~\eqref{Schroudinger equation} with the Gaussian coupling pulse
\eqref{eq:Omega-Gaussian} and the detuning profile \eqref{eq:Delta-sign-jump}.}
\label{fig2}
\end{figure}

Figure~\ref{fig1} shows final transition probability $P_{2}(\infty)$ as a function of the peak Rabi
frequency $\Omega_{0}$ for a fixed detuning $\Delta_{0}=5/T$. The solid
curve gives the numerical result, while the dotted curve represents the
analytic prediction \eqref{eq:P2-final-frac}. The two curves are
indistinguishable on the scale of the plot, confirming the validity of the
adiabatic--sudden treatment for these parameters.

\begin{figure}[htb]
\includegraphics[width=\columnwidth]{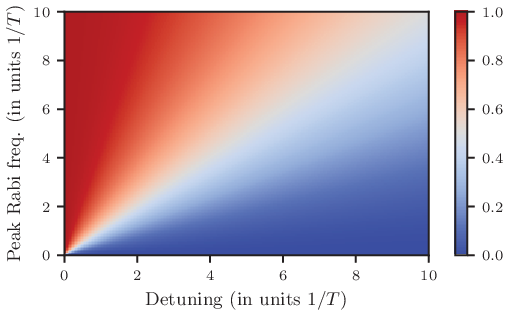}
\caption{(Color online) Contour plot of the final transition probability
$P_2(\infty)$ evaluated from the analytical expression
\eqref{eq:P2-final-frac}, using the same parameter ranges as in
Fig.~\ref{fig2}.}
\label{fig3}
\end{figure}

A more global comparison is presented in Figs.~\ref{fig2} and \ref{fig3},
which display contour plots of $P_{2}(\infty)$ in the
$(\Omega_{0},\Delta_{0})$ plane obtained from the numerical integration
(Fig.~\ref{fig2}) and from the analytic expression Eq.\eqref{eq:P2-final-frac} (Fig.~\ref{fig3}), respectively. A broad region of
high transition probability is clearly visible, as one would expect from
adiabatic passage. However, as discussed in Secs.~\ref{sec:Step adiabatic evolution and total propagator}, the underlying mechanism is a combination of adiabatic
evolution (before and after the jump) and a single sudden change of the
adiabatic basis at the detuning flip. The comparison between
Figs.~\ref{fig2} and \ref{fig3} shows excellent agreement over most of the
parameter space; noticeable deviations occur only in the near-resonant
region $\Delta_{0} \approx 0$, where the adiabaticity condition
\eqref{eq:adiabaticity-inequality} as well as asymptotics \eqref{eq:asymptotic1} and \eqref{eq:asymptotic2} are no longer well satisfied and the
adiabatic--sudden approximation breaks down. Indeed, in this nonadiabatic
regime the remaining discrepancies correspond to Rabi-like oscillations,
equivalent to those occurring outside the adiabatic-approximation regime.

Finally, we have verified numerically that the same conclusions hold for $\sech$- and Lorentzian-shaped coupling pulses: the resulting transition probabilities follow Eq.~\eqref{eq:P2-final-frac} with essentially the same accuracy as in the Gaussian case, and are therefore not shown here.

\section{Extension to SU(2)-symmetric three-state system}

\label{sec:lambda}

\subsection{Majorana Decomposition}

\begin{figure}[htb]
\includegraphics[width=0.5\columnwidth]{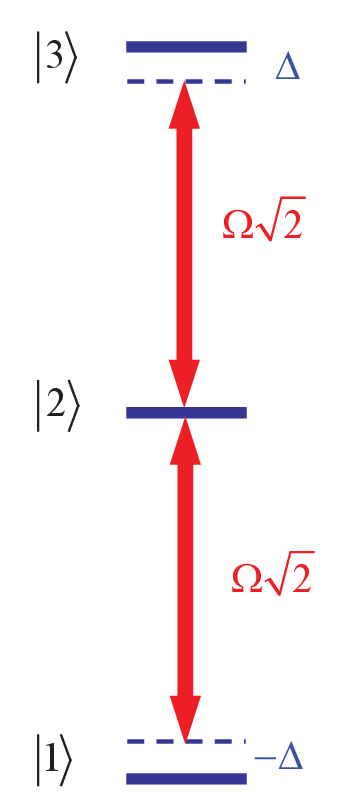}
\caption{(Color online) Schematic of a three-state, chainwise-coupled quantum system exhibiting SU(2) dynamical symmetry.
}
\label{fig4}
\end{figure}

We now consider the three-state system sketched in Fig.~\ref{fig4}, which is the
simplest nontrivial example of a multistate system with SU(2) dynamical
symmetry \cite{Majorana,Bloch,Hioe}. In the ordered basis
$\{\ket{1},\ket{2},\ket{3}\}$ the Hamiltonian in the rotating-wave
approximation )\cite{Allen-Eberly,Shore} can be written as
\begin{equation}
H_{3}(t)=\hbar
\begin{pmatrix}
-\Delta(t) & \Omega(t)/\sqrt{2} & 0 \\[4pt]
\Omega(t)/\sqrt{2} & 0 & \Omega(t)/\sqrt{2} \\[4pt]
0 & \Omega(t)/\sqrt{2} & \Delta(t)
\end{pmatrix},
\label{eq:H-3level-Majorana}
\end{equation}
where $\Omega(t)$ and $\Delta(t)$ are exactly the same coupling and detuning
functions as in the two-level model of Sec.~\ref{sec:twolevel}.

Because all couplings are proportional to the same $\Omega(t)$ and
the diagonal elements form a simple ``ladder'' in $\Delta(t)$, the dynamics
of this chain is completely determined by the \emph{same} SU(2) rotation
that governs the two-state system. In Majorana’s construction
\cite{Majorana,Genov}, the three-state dynamics can be obtained by
embedding the original two-level problem into the symmetric subspace of two
fictitious spin-$\tfrac12$ systems. As a consequence, the three-state
propagator is not independent: it is fully expressed in terms of the
Cayley--Klein parameters $a$ and $b$ that describe the two-level evolution
[Eq.~\eqref{U}].

Denoting the three-state propagator by $U_{3}(t_{f},t_{i})$, one finds
\cite{Genov}
\begin{equation}
U_{3}(t_{f},t_{i}) =
\begin{pmatrix}
a^{2} & \sqrt{2}\,a b & b^{2} \\[4pt]
-\sqrt{2}\,a b^{\ast} & |a|^{2}-|b|^{2} & \sqrt{2}\,a^{\ast} b \\[4pt]
-b^{\ast 2} & -\sqrt{2}\,a^{\ast} b^{\ast} & a^{\ast 2}
\end{pmatrix},
\label{eq:U3-Majorana}
\end{equation}
with the unitarity condition $|a|^{2}+|b|^{2}=1$ inherited from the
underlying two-level propagator. In our detuning-jump protocol the
parameters $a$ and $b$ are given explicitly by
Eqs.~\eqref{eq:a-CK}–\eqref{eq:b-CK}; substituting them into
Eq.~\eqref{eq:U3-Majorana} provides the full three-state evolution matrix
for the SU(2)-symmetric chain driven by the detuning sign jump.

We now use Eq.~\eqref{eq:U3-Majorana} to evaluate the transition
probabilities for different initial conditions. For brevity we define
\begin{equation}
s = \sin\delta\theta, \qquad c = \cos\delta\theta,
\end{equation}
so that, with Eqs.~\eqref{eq:a-CK}–\eqref{eq:b-CK},
\begin{equation}
|a|^{2} = s^{2}, \qquad |b|^{2} = c^{2}.
\end{equation}

\subsection{Initial state \texorpdfstring{$\ket{1}$}{|1⟩}}

If the system is initially prepared in $\ket{1}$, the initial state vector
is $(1,0,0)^{T}$, and the final state is given by the first column of
$U_{3}(t_{f},t_{i})$,
\begin{equation}
\ket{\psi_{f}} =
\begin{pmatrix}
a^{2} \\[3pt]
- \sqrt{2}\,a b^{\ast} \\[3pt]
- b^{\ast 2}
\end{pmatrix}.
\end{equation}
The transition probabilities to the three states are therefore
\begin{align}
P_{1\to 1} &= |a^{2}|^{2} = |a|^{4} = s^{4}, \\[4pt]
P_{1\to 2} &= 2\,|a b^{\ast}|^{2} = 2\,|a|^{2} |b|^{2} = 2 s^{2} c^{2}, \\[4pt]
P_{1\to 3} &= |b^{\ast 2}|^{2} = |b|^{4} = c^{4}.
\end{align}
As in the two-level case, the probabilities depend only on the mixing-angle
jump $\delta\theta$ through $s$ and $c$, and are independent of the
dynamical phases $\delta_{\pm}$.

In our symmetric detuning-jump case with $\Delta(t)=\pm\Delta_{0}$ and peak
coupling $\Omega_{0}$ one has
\begin{equation}
s^{2} = \frac{\Delta_{0}^{2}}{\Omega_{0}^{2} + \Delta_{0}^{2}},
\qquad
c^{2} = \frac{\Omega_{0}^{2}}{\Omega_{0}^{2} + \Delta_{0}^{2}},
\end{equation}
so that
\begin{align}
P_{1\to 1} &= \left( \frac{\Delta_{0}^{2}}
                         {\Omega_{0}^{2} + \Delta_{0}^{2}} \right)^{2}, \\[4pt]
P_{1\to 2} &= \frac{2\,\Omega_{0}^{2}\Delta_{0}^{2}}
                   {(\Omega_{0}^{2} + \Delta_{0}^{2})^{2}}, \\[4pt]
P_{1\to 3} &= \left( \frac{\Omega_{0}^{2}}
                         {\Omega_{0}^{2} + \Delta_{0}^{2}} \right)^{2}.
\end{align}
In the strong-coupling regime $\Omega_{0} \gg \Delta_{0}$, we obtain
\begin{equation}
P_{1\to 1} \approx 0, \qquad
P_{1\to 2} \approx 2\left(\frac{\Delta_{0}}{\Omega_{0}}\right)^{2}, \qquad
P_{1\to 3} \approx 1 - 2\left(\frac{\Delta_{0}}{\Omega_{0}}\right)^{2},
\end{equation}
i.e.\ nearly complete transfer from $\ket{1}$ to $\ket{3}$, with a small
leakage into $\ket{2}$.

\subsection{Initial state \texorpdfstring{$\ket{2}$}{|2⟩}}

If the system starts in the middle state $\ket{2}$, the initial vector is
$(0,1,0)^{T}$ and the final state is the second column of $U_{3}$,
\begin{equation}
\ket{\psi_{f}} =
\begin{pmatrix}
\sqrt{2}\,a b \\[3pt]
|a|^{2} - |b|^{2} \\[3pt]
- \sqrt{2}\,a^{\ast} b^{\ast}
\end{pmatrix}.
\end{equation}
The corresponding probabilities are
\begin{align}
P_{2\to 1} &= 2\,|a b|^{2} = 2 s^{2} c^{2}, \\[4pt]
P_{2\to 2} &= (|a|^{2} - |b|^{2})^{2} = (s^{2} - c^{2})^{2}, \\[4pt]
P_{2\to 3} &= 2\,|a^{\ast} b^{\ast}|^{2} = 2 s^{2} c^{2}.
\end{align}
Thus the population leaks symmetrically from $\ket{2}$ into $\ket{1}$ and
$\ket{3}$, while the remaining probability in the middle state is governed
by the difference $s^{2} - c^{2}$.

For our symmetric detuning-jump protocol this becomes
\begin{align}
P_{2\to 1} &= P_{2\to 3}
 = \frac{2\,\Omega_{0}^{2}\Delta_{0}^{2}}{(\Omega_{0}^{2} + \Delta_{0}^{2})^{2}},
\\[4pt]
P_{2\to 2} &= \left(
\frac{\Delta_{0}^{2} - \Omega_{0}^{2}}{\Omega_{0}^{2} + \Delta_{0}^{2}}
\right)^{2}.
\end{align}
In the limit $\Omega_{0} \gg \Delta_{0}$ this yields
\begin{equation}
P_{2\to 1} \approx P_{2\to 3} \approx
2\left(\frac{\Delta_{0}}{\Omega_{0}}\right)^{2}, \qquad
P_{2\to 2} \approx 1 - 4\left(\frac{\Delta_{0}}{\Omega_{0}}\right)^{2},
\end{equation}
showing that the middle state is only weakly affected by the detuning-jump
protocol when the coupling is large.

\subsection{Initial state \texorpdfstring{$\ket{3}$}{|3⟩}}

Finally, if the system is initially in $\ket{3}$, the initial vector is
$(0,0,1)^{T}$ and the final state is the third column of $U_{3}$,
\begin{equation}
\ket{\psi_{f}} =
\begin{pmatrix}
b^{2} \\[3pt]
\sqrt{2}\,a^{\ast} b \\[3pt]
a^{\ast 2}
\end{pmatrix}.
\end{equation}
The transition probabilities are
\begin{align}
P_{3\to 1} &= |b^{2}|^{2} = |b|^{4} = c^{4}, \\[4pt]
P_{3\to 2} &= 2\,|a^{\ast} b|^{2} = 2 s^{2} c^{2}, \\[4pt]
P_{3\to 3} &= |a^{\ast 2}|^{2} = |a|^{4} = s^{4}.
\end{align}
Thus the pattern is the mirror image of the case with initial $\ket{1}$.

For the symmetric detuning-jump case this gives
\begin{align}
P_{3\to 1} &= \left( \frac{\Omega_{0}^{2}}
                         {\Omega_{0}^{2} + \Delta_{0}^{2}} \right)^{2}, \\[4pt]
P_{3\to 2} &= \frac{2\,\Omega_{0}^{2}\Delta_{0}^{2}}
                   {(\Omega_{0}^{2} + \Delta_{0}^{2})^{2}}, \\[4pt]
P_{3\to 3} &= \left( \frac{\Delta_{0}^{2}}
                         {\Omega_{0}^{2} + \Delta_{0}^{2}} \right)^{2},
\end{align}
and in the strong-coupling limit $\Omega_{0} \gg \Delta_{0}$,
\begin{equation}
P_{3\to 1} \approx 1 - 2\left(\frac{\Delta_{0}}{\Omega_{0}}\right)^{2}, \qquad
P_{3\to 2} \approx 2\left(\frac{\Delta_{0}}{\Omega_{0}}\right)^{2}, \qquad
P_{3\to 3} \approx 0,
\end{equation}
corresponding to almost complete population transfer from $\ket{3}$ to
$\ket{1}$.

\section{Relation to piecewise adiabatic evolution}

The present detuning--sign--jump protocol belongs to the broader family of
robust schemes that combine adiabatic ideas with short, localized
nonadiabatic events. It is instructive to compare our approach with the
various implementations of piecewise adiabatic passage (PAP) developed in
Refs.~\cite{ShapiroPAP1,ZhdanovichPAP,ShapiroPAP2}. In PAP, a conventional
adiabatic process (such as frequency-chirped rapid adiabatic passage in a
two-level system or STIRAP \cite{Vitanov2017} in a three-level system) is first chosen as a
continuous-time ``reference'' evolution. This smooth evolution is then
partitioned into many short time intervals, during each of which only a small
fraction of the population is allowed to move between the instantaneous
eigenstates. Within each interval, the smooth driving fields are replaced by
a short ``kick'' (typically a femtosecond pulse \cite{Wollenhaupt,Brixner} or a pulse pair), chosen such
that the \emph{integrated} action of the field on that interval reproduces the
reference adiabatic dynamics. The overall transfer is achieved by coherent
accumulation of the effect of many such kicks, with robustness inherited from
the underlying adiabatic reference.

Our scheme shares with PAP the key conceptual ingredient of
\emph{stepwise adiabatic evolution}: for most of the time the system follows
an instantaneous adiabatic eigenstate, and the population transfer is driven
by brief departures from strict adiabaticity. However, there are also several
important differences. First, PAP relies on a \emph{large number} of weakly
nonadiabatic kicks (one per pulse or pulse pair), separated by intervals of
free evolution. The adiabatic parameters (effective Rabi frequencies and
detunings, or mixing angle) vary slowly \emph{on average}, but the instantaneous
field is strongly modulated from pulse to pulse. By contrast, in our
detuning--jump protocol the coupling $\Omega(t)$ is a single smooth pulse and
the nonadiabaticity is concentrated in \emph{one} well-defined event: the
sudden sign flip of the detuning. In the adiabatic basis this produces a
single localized ``kick'' with strength controlled by the mixing-angle jump
$\delta\theta$, while the evolution for $t<0$ and $t>0$ is strictly adiabatic.
This simplification allows us to derive the exact two-state propagator in
terms of just three parameters ($\delta\theta$ and the dynamical phases
$\delta_{\pm}$) and to express the final transition probability in the compact
form~\eqref{eq:P2-final-cos2}.

Second, in PAP the control knobs are the amplitudes, phases, and timing of a
train of pulses: the adiabaticity condition is enforced at the level of a
coarse-grained envelope of the pulse train, while the underlying field may be
highly structured in time. In our scheme the field amplitude is essentially
featureless apart from its smooth envelope, and the only nontrivial shaping
is a binary sign change of the detuning. The robustness of the protocol is
therefore tied directly to the geometric quantity $\delta\theta$, determined
by the ratio $\Omega_0/\Delta_0$ at the jump, rather than to the detailed
structure of a pulse train. This physical simplicity makes it particularly
transparent how parameter variations affect the final populations and how the
protocol interpolates between adiabatic and strongly nonadiabatic regimes.

\section{Possible implementations}
\label{sec:implementation}

The detuning--sign--jump protocol requires only three basic ingredients:
(i) a two-state (or SU(2)-symmetric multistate) manifold,
(ii) the ability to apply a smooth pulse envelope for the coupling $\Omega(t)$,
and (iii) a way to invert the sign of the detuning $\Delta(t)$ on a time
scale short compared to $1/\varepsilon(0)$, while keeping $\Omega(t)$
essentially unchanged across the jump. These requirements are modest and can
be met in several well-established experimental platforms.

\subsection{Two-level spin and microwave implementations}

Perhaps the most straightforward realization is a driven spin-$\tfrac12$
system, such as a nuclear or electronic spin in NMR/ESR, or a hyperfine/Zeeman
qubit in trapped ions and neutral atoms. In such systems the two levels form
an excellent isolated two-state manifold, and both the amplitude and the
frequency of the driving microwave or rf field are routinely shaped using
arbitrary waveform generators.

The smooth Rabi envelope $\Omega(t)$ is implemented by standard amplitude
modulation of the drive. The detuning $\Delta(t)$ is simply the difference
between the drive frequency and the qubit transition frequency; a sign flip
$\Delta_0 \to -\Delta_0$ can be realized by a rapid step change of the drive
frequency across resonance, or by a small step change of the static magnetic
field. The relevant time scales (typically microseconds or longer) make it
realistic to implement a detuning jump that is effectively sudden on the
scale of $1/\varepsilon(0)$, yet technically well controlled. The same
hardware has already been used to implement phase-jump and composite-pulse
protocols \cite{Vitanov2007,Torosov2007,Lehto,Zlatanov}, so the present
scheme fits naturally into existing control toolboxes.

\subsection{Optical and molecular realizations}

The protocol can also be implemented on near-resonant optical transitions in
atoms or molecules. A smooth $\Omega(t)$ can be generated by shaping the
intensity envelope of a single laser pulse using an acousto-optic or
electro-optic modulator, or a Fourier-domain pulse shaper. A detuning sign
flip may then be obtained, for example, by switching between two optical
fields that share the same smooth envelope but have opposite detunings
$\pm\Delta_0$ from the transition, with the switch occurring near the pulse
maximum. In the ideal limit the combined intensity envelope remains smooth,
while the effective detuning follows the desired step profile
\eqref{eq:Delta-sign-jump}.

More sophisticated optical arrangements, such as frequency-comb-based
control of molecular levels, have already been used to realize piecewise
adiabatic passage schemes \cite{ShapiroPAP1,ZhdanovichPAP,ShapiroPAP2}. In
such setups, the same pulse-shaping infrastructure could, in principle, be
employed to engineer a single smooth pulse with a controlled detuning jump,
providing an alternative to pulse-train-based piecewise adiabatic protocols.

\subsection{SU(2)-symmetric three-state chains}

For the SU(2)-symmetric three-state chain of Sec.~\ref{sec:lambda}, the
implementation requirements are essentially those of the underlying
two-level problem, together with an internal structure that realizes the
spin-1 coupling pattern. A concrete example is provided by three Zeeman
sublevels $m=-1,0,+1$ of a hyperfine manifold in a static magnetic field,
driven by a linearly polarized rf or microwave field, as discussed in
Refs.~\cite{Hioe,Genov}. With a suitable choice of geometry and
polarization, the nearest-neighbor couplings acquire the $\Omega(t)/\sqrt{2}$
structure of Eq.~\eqref{eq:H-3level-Majorana}, and the common detuning
$\Delta(t)$ is again set by the difference between the drive frequency and
the Zeeman splitting.

In this case, the same detuning jump used for the effective spin-$\tfrac12$
system (a fast step of the drive frequency or magnetic field) automatically
implements the three-state protocol. The full three-state propagator then
follows from the two-level Cayley--Klein parameters via
Eq.~\eqref{eq:U3-Majorana}, ensuring that any robust inversion achieved in
the two-level setting is mirrored by robust transfer between the outer
states $\ket{1}$ and $\ket{3}$ in the three-state chain.

\section{Conclusions}

\label{sec:conclusions}

We have proposed a simple protocol for robust population inversion based on
a detuning sign jump at the maximum of a smooth coupling pulse. In a
two-level system with a smooth pulse-shape coupling (Gaussian, $\sech$,
Lorentzian, etc.) and a detuning profile of constant magnitude but opposite
signs before and after the pulse center, the dynamics can be understood as
adiabatic evolution interrupted by a single nonadiabatic event in the
adiabatic basis. Within an adiabatic--sudden approximation we derived the
compact expression~\eqref{eq:P2-final-cos2} for the final excited-state
population, showing that high-fidelity inversion is achieved whenever the
peak coupling exceeds the detuning. Numerical simulations can be used to
confirm the robustness of the protocol and to delineate the regime where the
approximations hold.

We then exploited the Majorana construction to extend the same idea to an
SU(2)-symmetric three-state chain driven by the \emph{same} coupling and
detuning functions. In this case the three-state propagator is obtained
directly from the two-level Cayley--Klein parameters via
Eq.~\eqref{eq:U3-Majorana}, so that the entire three-state dynamics is fixed
by the same detuning-jump protocol. This mapping allowed us to derive
closed-form transition probabilities for all three initial states in terms
of the mixing-angle jump $\delta\theta$, demonstrating, in particular,
near-complete population transfer between the two outer states in the
strong-coupling regime, with only small population in the middle
state.

Because the three-state model is an SU(2) embedding of the original
two-level problem, all robustness properties of the detuning-jump protocol
carry over automatically: moderate variations of pulse area, detuning, or
exact timing of the jump leave the final populations largely unaffected as
long as the stepwise adiabaticity and sudden-jump conditions are
satisfied. This suggests that detuning-sign-jump control in SU(2)-symmetric
chains, and more generally in higher-spin realizations of two-level
dynamics, can provide a practical route to implementing simple, analytically
tractable, and robust population-transfer schemes in multistate quantum
systems.

\section*{Acknowledgements}

This research is partially supported by the Bulgarian national plan for
recovery and resilience, contract BG-RRP-2.004-0008-C01 SUMMIT: Sofia
University Marking Momentum for Innovation and Technological Transfer,
project number 3.1.4.



\begin{thebibliography}{99}
\bibitem{Allen-Eberly} L. Allen and J. H. Eberly, \emph{Optical Resonance
and Two-Level Atoms} (Wiley, New York, 1975).

\bibitem{Shore} B. W. Shore, \emph{The Theory of Coherent Atomic Excitation}
(Wiley, New York, 1990).

\bibitem{Yatsenko} L. P. Yatsenko, B. W. Shore, T. Halfmann, K. Bergmann,
and A. Vardi, Phys. Rev. A \textbf{60}, R4237(R) (1999).

\bibitem{Vitanov2001A} N.V. Vitanov, T. Halfmann, B.W. Shore, and K.
Bergmann, Annu. Rev. Phys. Chem. \textbf{52}, 763 (2001).

\bibitem{Vitanov2001B} N.V. Vitanov, M. Fleischauer, B. W. Shore and K.
Bergmann, Adv. At. Mol. Opt. Phys. \textbf{46}, 55 (2001).

\bibitem{Klein} J. Klein, F. Beil, and T. Halfmann, J. Phys. B \textbf{40},
S345 (2007).

\bibitem{Levitt79} M. H. Levitt and R. Freeman, J. Magn. Reson. \textbf{33},
473 (1979).

\bibitem{Freeman80} R. Freeman, S. P. Kempsell, and M. H. Levitt, J. Magn.
Reson. \textbf{38}, 453 (1980).

\bibitem{Levitt86} M. H. Levitt, Prog. Nucl. Magn. Reson. Spectrosc. \textbf{%
18}, 61 (1986).

\bibitem{Freeman97} R. Freeman, \emph{Spin Choreography} (Spektrum, Oxford,
1997).

\bibitem{Vitanov2007} N. V. Vitanov, New J. Phys. \textbf{9}, 58 (2007).

\bibitem{Torosov2007} B. T. Torosov and N. V. Vitanov, Phys. Rev. A \textbf{%
76}, 053404 (2007).

\bibitem{Lehto} J. M. S. Lehto and K.-A. Suominen, Phys. Rev. A \textbf{94},
013404 (2016)

\bibitem{Zlatanov} K. N. Zlatanov and  N. V. Vitanov, Phys. Rev. A \textbf{101}, 013426 (2020).

\bibitem{Majorana} E. Majorana, Nuovo Cimento \textbf{9}, 43 (1932).

\bibitem{Bloch} F. Bloch and I. I. Rabi, Rev. Mod. Phys. \textbf{17}, 237 (1945).

\bibitem{Hioe} F. T. Hioe, J. Opt. Soc. Am. B \textbf{4}, 1327 (1987).

\bibitem{Ramakrishna} V. Ramakrishna, K. Flores, H. Rabitz, and R. Ober, Phys. Rev. A \textbf{62}, 053409 (2000).


\bibitem{Genov} G. T. Genov, B. T. Torosov, and N. V. Vitanov, Phys. Rev. A \textbf{84}, 063413 (2011).

\bibitem{ShapiroPAP1} E. A. Shapiro, V. Milner, C. Menzel-Jones, and M. Shapiro,
Phys. Rev. Lett. \textbf{99}, 033002 (2007).

\bibitem{ZhdanovichPAP} S. Zhdanovich, E. A. Shapiro, M. Shapiro, J. W. Hepburn, and V. Milner,
Phys. Rev. Lett. \textbf{100}, 103004 (2008).

\bibitem{ShapiroPAP2} E. A. Shapiro, A. Pe'er, J. Ye, and M. Shapiro,
Phys. Rev. Lett. \textbf{101}, 023601 (2008).

\bibitem{Vitanov2017} N. V. Vitanov, A. A. Rangelov, B. W. Shore, and K. Bergmann, Rev. Mod. Phys. \textbf{89}, 015006 (2017).

\bibitem{Wollenhaupt} M. Wollenhaupt, V. Engel, and T. Baumert, Annu. Rev. Phys. Chem. \textbf{56}, 25 (2005).

\bibitem{Brixner} T. Brixner, T. Pfeifer, G. Gerber, M. Wollenhaupt, and T. Baumert, in \emph{Femtosecond Laser Spectroscopy}, ed. P. Hannaford
(Springer, New York, 2005), Chap. 9.

\end{thebibliography}
\end{document}